\documentclass[,final]{aipproc}

\layoutstyle{6x9}

\begin{document}

\title{ Measurements of Transverse Spin Effects with the Forward Pion Detector of STAR }


\author{ L. Nogach for the STAR collaboration }{
  address={Institute of High Energy Physics \\
  1 Pobeda street, Protvino, Moscow region 142281, Russia}
}

\begin{abstract}
Measurements by the STAR collaboration of neutral pion production 
at large Feynman x ($x_F$) in the first polarized proton collisions at 
$\sqrt{s}=200$~GeV were reported previously. Cross sections measured 
at $\eta=3.3$, 3.8 and 4.0 are found to be consistent with next-to-leading 
order perturbative QCD calculations.  The analyzing power is consistent 
with zero at negative $x_F$ and at positive $x_F$ up to ~0.3, then grows 
more positive with increasing $x_F$. This behavior can be described by 
phenomenological models including the Sivers effect, the Collins effect 
or higher twist contributions in the initial and final states. 
Forward calorimetry at STAR has been extended, and there are plans for
further expansion.  An integrated luminosity of 6.8 pb$^{-1}$ with average 
beam polarization of 60\% from online polarimetry measurements was sampled 
with the upgraded FPD in the 2006 RHIC run. This data sample will allow 
for a detailed map of the $\pi^0$ analyzing power over kinematic variables 
bounded by $0.3 < x_F < 0.6$ and $1.2 < p_T < 5.0$ GeV/$c$ at 
$\sqrt{s}=200$~GeV.  
The expanded FPD has observed multi-photon final states expected to 
have "jet-like" characteristics.  The transverse spin dependence of 
jet-like events can discriminate between the Collins and Sivers effects 
and lead to further progress in understanding the origin of single spin 
asymmetries in forward particle production.
Data were also obtained at $\sqrt{s}=62.4$~GeV for $x_F\rightarrow 1$ 
to test predictions based on phenomenological fits to earlier STAR results.  
Recent results, the status of the analysis of 2006 run data and near-term 
plans will be discussed.
\end{abstract}

\maketitle

\vspace*{0.5cm}
In about twenty years after the first measurements of significant spin effects
in hadronic reactions \cite{hyp_pol,e704_pi}, the study of single spin asymmetry 
$A_N$ for $\pi^0$ inclusive production is still of current interest. 
As is known, large single spin effects cannot be explained within 
collinear perturbative QCD at leading twist due to helicity conservation. A number 
of theoretical models based on generalizations of the factorization theorem 
were proposed to account for significant values of $A_N$. These models 
assume the presence of (i) higher twist correlation functions in the initial 
or final state (for example, twist-3 in the Qui-Sterman \cite{QS} and 
Efremov-Teryaev \cite{ET} models), (ii) parton intrinsic transverse 
momentum $k_T$ and spin dependence of the distribution functions (Sivers 
effect \cite{siv}), (iii) $k_T$ and spin dependence of the fragmentation 
functions (Collins effect \cite{col}). 
There is some indication from recent theoretical studies \cite{ansl} 
that with all partonic motion properly taken into account the Collins 
mechanism is suppressed and cannot alone explain large measured $A_N$. 
Consequently, both the Sivers and the Collins effects can contribute to 
the pion asymmetry, as they are observed to do so in semi-inclusive deep 
inelastic scattering from a transversely polarized target \cite{HERMES}. 

First measurements by STAR \cite{fpd_prl} showed that the $\pi^0$ $A_N$
revealed at lower energies persists for $pp$ collisions at $\sqrt{s}=200$~GeV 
and that the measured $A_N$ can be described by several models. 
Thus, more precise measurements were needed to distinguish between 
different mechanisms of the emergence of $A_N$. Owing to improvements 
in both luminosity and beam polarization at RHIC, an integrated luminosity of
6.8 pb$^{-1}$ with average beam polarization of 60\% was sampled
with the STAR Forward Pion Detector (FPD) in the 2006 run. 

The FPD are calorimeters consisting of lead glass cells located
on both sides of the STAR interaction region (IR) at a distance of about 
8~m (East FPD) and 7~m (West FPD++) and close to the beam pipe. They provide
for triggering and reconstruction of $\pi^0$-mesons at forward rapidities. 
The FPD++ was installed for the 2006 run as a prototype of the Forward 
Meson Spectrometer \cite{lcb_hp2004}.
The calorimeter modules on the two sides of the IR were placed
at different distances with respect to the beam line, corresponding
to the average values of pseudorapidity 3.7 and 3.3. This allowed to expand 
the covered range of transverse momentum at fixed $x_F$. A strong correlation
between $x_F$ and $p_T$ is observed in individual calorimeters because of 
their narrow acceptance.
The reconstruction algorithm, which uses a fit to the experimentally measured 
shower shape, was carefully studied previously with Monte-Carlo simulations. 
The energy scale of the detector was determined from the $\pi^0$ peak 
position in the di-photon invariant mass distribution. The accuracy of 
calibration was at the level of 2\%. The cross-ratio method is used 
to obtain $A_N$, thereby eliminating many possible sources 
of systematic errors in the measurements. 

\begin{figure}
  \includegraphics[height=.35\textheight]{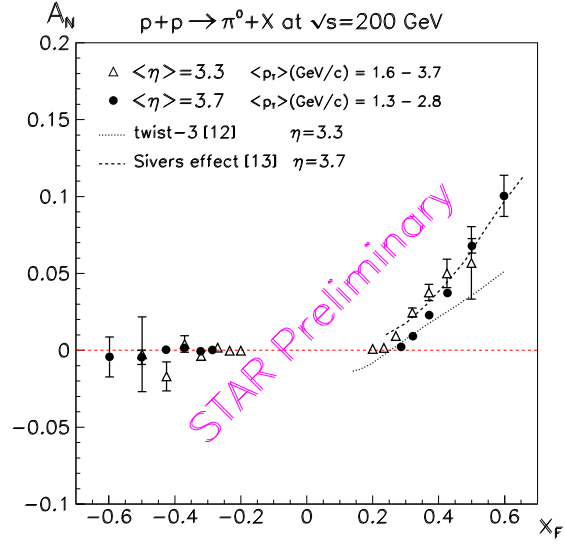}
  \caption{Analyzing power of $p+p\rightarrow\pi^0+X$ as a function of $x_F$
           at $<\eta>=3.3$ (triangles) and $<\eta>=3.7$ (circles). Error
           bars represent the statistical uncertainty. Systematic uncertainties
           not including a normalization error from the beam polarization 
           measurements are smaller than statistical ones.
           The lines are theoretical calculations.}
  \label{an-xf}
\end{figure}

The new measurements of $A_N$ as a function of $x_F$ are shown in Fig.\ref{an-xf}. 
Confirming the earlier measurements \cite{fpd_dubna05}, the analyzing 
power at positive $x_F$, relative to the polarized beam, grows from 
0 at $x_F\sim 0.2-0.3$ up to 0.1 at $x_F\sim 0.6$, and the $A_N$ 
at negative $x_F$, equivalent to the unpolarized beam direction,
is consistent with zero. Calculations from two theoretical models 
based on twist-3 contribution \cite{WV} and the Sivers effect \cite{UdA}
are also plotted. The high precision of these measurements allows for 
a quantitative comparison with theory predictions and should enable 
a discrimination between different dynamics.

\begin{figure}
  \includegraphics[height=.25\textheight]{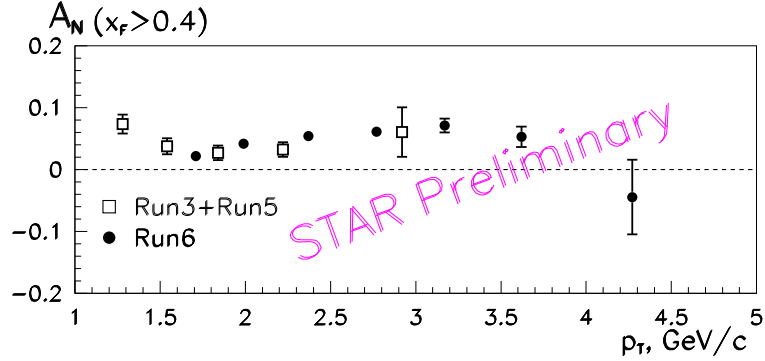}
  \caption{Dependence of $\pi^0$ $A_N$ on $p_T$ for $x_F>0.4$ from
           the 2006 run (circles) in comparison with the previous 
           measurements by STAR (squares).}
  \label{an-ptxfgt04}
\end{figure}

Fig.\ref{an-ptxfgt04} shows the analyzing power as a function of $p_T$ 
averaged on $x_F>0.4$ where the $A_N$ is significantly non-zero.
There was a hint from the previous measurements that $A_N$ decreases 
with increasing $p_T$ in the range $1\le p_T \le2$~GeV/$c$. 
In the 2006 run, the region of higher $p_T$ has been explored. 
The data in the overlapping region are consistent, but the 
$p_T$-dependence of the asymmetry looks more complicated now, 
although a sign in Fig.\ref{an-ptxfgt04} of the asymmetry growing 
at $p_T\sim 2-3$~GeV/$c$ may be a consequence of residual $x_F$-dependence. 

\begin{figure}
  \includegraphics[height=.48\textheight]{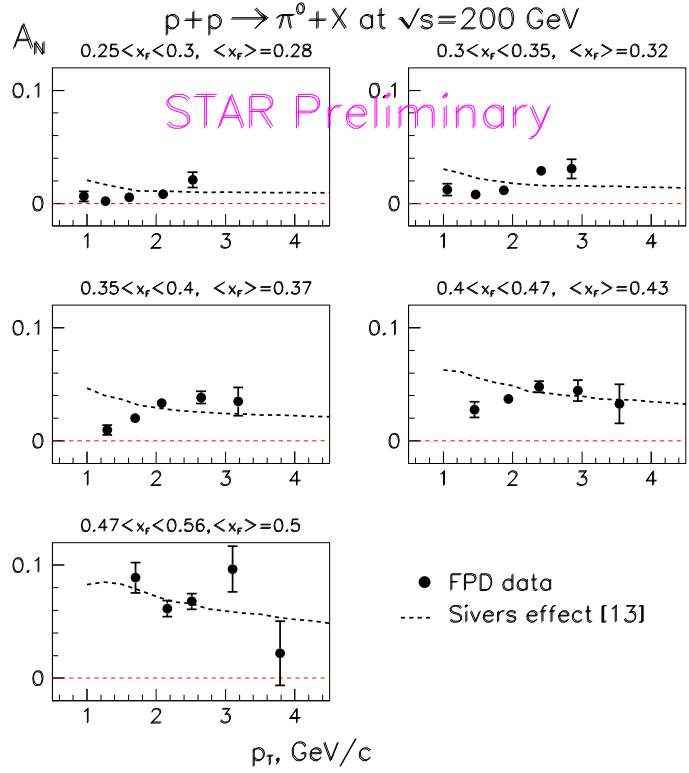}
  \caption{$\pi^0$ $A_N$ as a function of transverse momentum in $x_F$ bins.}
  \label{an-pt}
\end{figure}

To get rid of the correlation between the average values of $x_F$ and 
$p_T$ in the $p_T$ bins, the $p_T$ dependence of $A_N$ has been measured 
in five narrow $x_F$ intervals. The results based on the combined data 
from the three runs are presented in Fig.\ref{an-pt}.
An increase of $A_N$ with increasing $p_T$ is observed in all 
but the highest $x_F$ bin, where the statistics are limited. It should be 
pointed out that the unpolarized cross section for $\pi^0$ production 
in $pp$ collisions at $\sqrt{s}=200$~GeV is found to be generally consistent 
with next-to-leading order pQCD calculations \cite{cs_phenix,fpd_prl2} 
in this range of $x_F$ and $p_T$, unlike at lower energies.

All available models that aim to explain the $A_N$ for pion production 
predict a monotonic decrease of $A_N$ with increasing $p_T$, approximately 
falling as 1/$p_T$.  As an example, the results of calculations based on 
the Sivers effect \cite{UdA} are shown in comparison to the data 
in Fig.\ref{an-pt}.  The dependence of the data on $p_T$ is not explained
by the calculation, with the possible exception of the highest $x_F$ bin 
where the statistical errors of the measurement are large.  The data do not 
show a monotonic decrease of $A_N$ with increasing~$p_T$.

Also in the last run data were obtained with the FPD in $pp$ collisions 
at $\sqrt{s}=62.4$~GeV. Data analysis is in progress with a view to test
theoretical predictions based on phenomenological fits to earlier STAR 
results \cite{WV}. Future studies with the FMS which is now under construction 
at STAR and expected to be operational in the 2007 RHIC run, will include, 
among other things, further measurements of spin effects in $pp$ collisions. 
The FMS will provide full azimuthal coverage at $2.5<\eta<4.0$ and broader 
acceptance in $(x_F,p_T)$ plane. Thus, it is a well suited tool to isolate 
the Sivers mechanism from the Collins effect by measuring analyzing power 
in $\pi^0-\pi^0$ and jet production. The most interesting results to be
expected from these measurements is an estimate of the quark orbital momentum 
contribution to the spin of nucleon.

\bibliographystyle{aipproc}

\end{document}